\documentclass[aps,prl,twocolumn,showpacs,superscriptaddress]{revtex4-1}
\usepackage{graphicx,amsmath,amssymb,amsfonts}
\usepackage[latin1]{inputenc}
\usepackage{array}
\usepackage{multirow}
\usepackage{float}
\usepackage{color}
\usepackage[normalem]{ulem}
\begin{document}

\title{Magnetic properties of van der Waals layered single crystals DyOBr and SmOCl}

\author{Feihao Pan}
\author{Daye Xu}
\author{Songnan Sun}
\author{Jiale Huang}
\author{Chenglin Shang}
\author{Bingxian Shi}
\author{Xuejuan Gui}
\affiliation{Laboratory for Neutron Scattering and Beijing Key Laboratory of Optoelectronic Functional Materials and MicroNano Devices, Department of Physics, Renmin University of China, Beijing 100872, China}
\affiliation{Key Laboratory of Quantum State Construction and Manipulation (Ministry of Education), Renmin University of China, Beijing, 100872, China}

\author{Jianfei Qin}
\author{Hongliang Wang}
\author{Lijie Hao}
\affiliation{China Institute of Atomic Energy, PO Box-275-30, Beijing 102413, China}

\author{Jinchen Wang}
\author{Juanjuan Liu}
\author{Hongxia Zhang}
\author{Peng Cheng}
\email[Corresponding author: ]{pcheng@ruc.edu.cn}
\affiliation{Laboratory for Neutron Scattering and Beijing Key Laboratory of Optoelectronic Functional Materials and MicroNano Devices, Department of Physics, Renmin University of China, Beijing 100872, China}
\affiliation{Key Laboratory of Quantum State Construction and Manipulation (Ministry of Education), Renmin University of China, Beijing, 100872, China}

\begin{abstract}
Two-dimensional van der Waals single crystals DyOBr and SmOCl have been grown by flux method and their anisotropic magnetic properties are reported. DyOBr orders antiferromagnetically at T$_{N}$=9.5~K with magnetic moments lying along $a$-axis, similar as DyOCl. Its magnetic susceptibility shows an anomaly at T$^{*}$=30~K possibly due to the crystal field effect. Furthermore a 1/3 magnetization plateau is clearly observed under H$\parallel$a and H$\parallel$[110], which might be a field-induced spin-flop phase or some exotic quantum magnetic state. On the other hand, isostructural SmOCl undergoes an antiferromagnetic transition at T$_{N}$=7.1~K and exhibits a contrasting Ising-like perpendicular $c$-axis magnetic anisotropy, which could be well explained by our crystal field calculations. Both DyOBr and SmOCl are insulators with band gap of $\sim$5~eV, our results suggest they are promising in building van der Waals heterostructures and applications in multifunctional devices.

\end{abstract}

\maketitle

\section{Introduction}
Two-dimensional (2D) van der Waals (vdW) materials with intrinsic long-range magnetic order have received considerable research interest in recent years\cite{Burch2018,review1,reveiw2,CPB}. These cleavable crystals may retain magnetic ordering in few layer thickness when magnetic anisotropy is strong enough to resist the thermal fluctuations. Therefore they provide an ideal platform for developing novel nanoscale magnetic devices as well as studying the fundamental magnetism models in 2D limit. Besides, they are also capable to realize various exciting quantum and topological phases\cite{AHE,Sky1,2018Kondo,CrI3Neutron,MnBiTe,Fe3Co,PCheng_APL2020}. The current magnetic vdW material family is still far from meeting the demand of either research interest or practical applications. Exploring more vdW magnetic compounds is expected to yield an enriched landscape of emergent quantum phenomena and applications.

Ternary compound with transition/rare earth metal element, chalcogen and halogen element contains considerable magnetic vdW materials, which have shown promising properties for spintronic applications. Among them, bulk CrOCl is an antiferromagnet below T$_N$=13.6~K\cite{CrOCl1}. The vdW heterostructures based 2D CrOCl have been reported to exhibit tunable unconventional insulating state\cite{CrOCl2} and exchange bias effect\cite{CrOCl3}. CrSBr is an antiferromagnetic semiconductor with T$_N$=132~K, which shows large negative magnetoresistance coupled with magnetic order\cite{CrSBr1,CrSBr2}. FeOCl has a complex spiral magnetic structure below T$_N$=92~K\cite{FeOCl1} and 2D FeOCl flakes possess strong in-plane optical and electrical anisotropy\cite{FeOCl2}. Previously, we have identified DyOCl as a rare-earth based 2D vdW material with A-type antiferromagnetic structure below T$_N$=10~K\cite{DyOCl}. It has large magnetic moment and strong uniaxial magnetic anisotropy along $a$-axis. On the other hand, YbOCl with the honeycomb lattice was recently identified without long-range magnetic order down to 1.8~K and proposed to be a Kitaev spin liquid candidate\cite{ZhangQM}. Besides, plenty of vdW members in this material family with long-range magnetic order have been unexplored, especially in single crystal form.   

In this paper, we report the synthesize of two magnetic vdW single crystals DyOBr and SmOCl, which are isostructural to DyOCl. The anisotropic magnetic properties of DyOBr are similar as DyOCl, which shows strong in-plane magnetic anisotropy, as well as field-induced magnetization plateau and spin-flip transition. By contrast, SmOCl exhibits perpendicular magnetic anisotropy and robust antiferromagnetism under field. The contrasting magnetic anisotropy can be well explained based our crystal field calculations and may stimulate further researches in designing nanoscale magneto-devices based on DyOBr and SmOCl crystals.

\begin{figure*}[htbp]
	\centering
	\includegraphics[width=\textwidth]{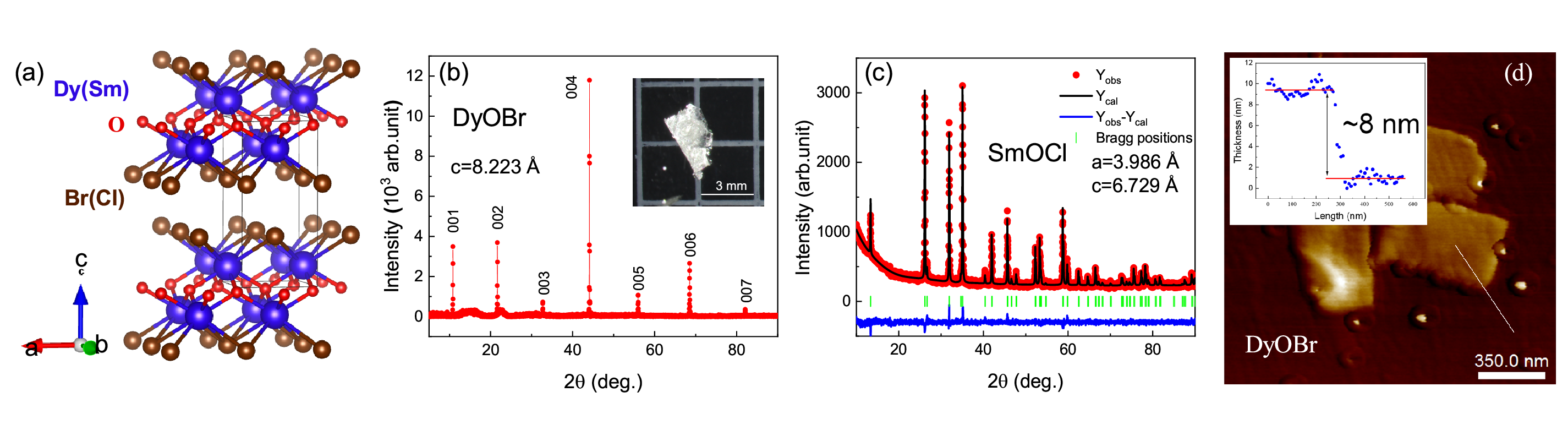}
	\caption {(a) Crystal structure of DyOBr. Isostructural SmOCl has relatively much smaller $c$-axis lattice parameter. (b) The x-ray reflections from the $ab$-plane of DyOBr single crystal. The inset shows the photo of one crystal. (c) Room temperature XRD patterns of SmOCl powders and Rietveld refinement result. (d) Atomic force microscopy images and height profile step of DyOBr nano-flakes mechanically-exfoliated onto a 300 nm SiO$_{2}$/Si substrate.} \label{Fig1}
\end{figure*}

\section{methods}

The polycrystalline samples of DyOBr and SmOCl were prepared by the solid state reaction. The raw materials, Dy$_{2}$O$_{3}$ and NH$_{4}$Br (or Sm$_{2}$O$_{3}$ and NH$_{4}$Cl), were mixed in a 1:3 mole ratio and pressed into a pellet. The pellet was then loaded into an alumina crucible and placed in a muffle furnace. The temperature was raised to $420\,^{\circ}\mathrm{C}$ ($450\,^{\circ}\mathrm{C}$ for SmOCl) and held for 1.5 hours and then $720\,^{\circ}\mathrm{C}$ ($650\,^{\circ}\mathrm{C}$ for SmOCl) for 1.5 hours. Then the product was furnace-cooled to room temperature.
DyOBr and SmOCl single crystals were grown by the flux method. The raw materials, Dy$_{2}$O$_{3}$ and DyBr$_{3}$ (or Sm$_{2}$O$_{3}$ and SmCl$_{3}$), were mixed in a 1:10 mole ratio and loaded into an alumina crucible. The crucible was sealed in a quartz tube under vacuum and filled with nearly one quarter atmosphere of pure argon gas. The tube was then placed in a muffle furnace and heated to $1150\,^{\circ}\mathrm{C}$ in 23 hours. The temperature was held at $1150\,^{\circ}\mathrm{C}$ for 12 hours and then cooled to $700\,^{\circ}\mathrm{C}$ at a rate of $1\,^{\circ}\mathrm{C}$ per hour. Transparent DyOBr single crystals with dimension up to $1.00 \times 0.50 \times 0.05~mm^{3}$ [inset of Fig. \ref{Fig1}(b)] could be obtained and excess flux can be dissolved by water. The typical size of SmOCl single crystal is smaller which is about $0.65 \times 0.35 \times 0.06~mm^{3}$. All the samples are air-stable and we have not found any evidence for the possible variation of the O or Cl content in different crystals.

X-ray diffraction (XRD) patterns of powder samples were collected from a Bruker D8 Advance X-ray diffractometer using Cu K$_{\alpha}$ radiation. Magnetization and heat capacity measurements were carried out in Quantum Design MPMS3 and PPMS-14T, respectively. The dimensions of exfoliated DyOBr and SmOCl nanoflakes were checked by a Bruker edge dimension atomic force microscope. The powder neutron diffraction experiments on DyOBr were carried out on Xingzhi cold neutron triple-axis spectrometer at the China Advanced Research Reactor (CARR)\cite{XingZhi}. The diffusion reflectance spectroscopy was measured on a Shimadzu UV-3600 UV-VIS-NIR spectrophotometer. The theoretical crystal electric field energy levels and corresponding wave functions are calculated using point charge model by McPhase software\cite{McPhase}.

\section{Results and discussions}

Both DyOBr and SmOCl adopt the tetragonal symmetry with space group $P4/nmm$ (No.129), same as DyOCl and previous reports\cite{DyOCl,DOB_cal,DyOBr_mag}. As shown in Fig. \ref{Fig1}(a), the rare earth atoms form square lattice within the $ab$-plane and the vdW gap exits between Br- or Cl-layers. As shown in Fig. 1 and the first two tables in the supplementary material, through the refinement from both powder and single crystal XRD patterns, the lattice parameter for DyOBr ($a=3.857$\text{\AA}, $c=8.224$\text{\AA} from powder and $c=8.223$\text{\AA} from single crystal) and SmOCl ($a=3.986$\text{\AA}, $c=6.729$\text{\AA}) could be obtained\cite{Supple}. One can see that the $c$-axis lattice constant for DyOBr is much larger than that of DyOCl and SmOCl, which might result in a larger vdW gap and make DyOBr a more two-dimensional compound. A recent first-principles calculation work has shown that single-layer DyOBr is stable and has cleavage energy comparable with CrI$_3$\cite{DOB_cal}. We have performed micro-mechanical exfoliation of DyOBr single crystals using Scotch tape and nanosheet with thickness of 8~nm could be easliy obtained as shown in Fig. \ref{Fig1}(d).

\begin{figure}
	\includegraphics[width=7.5cm]{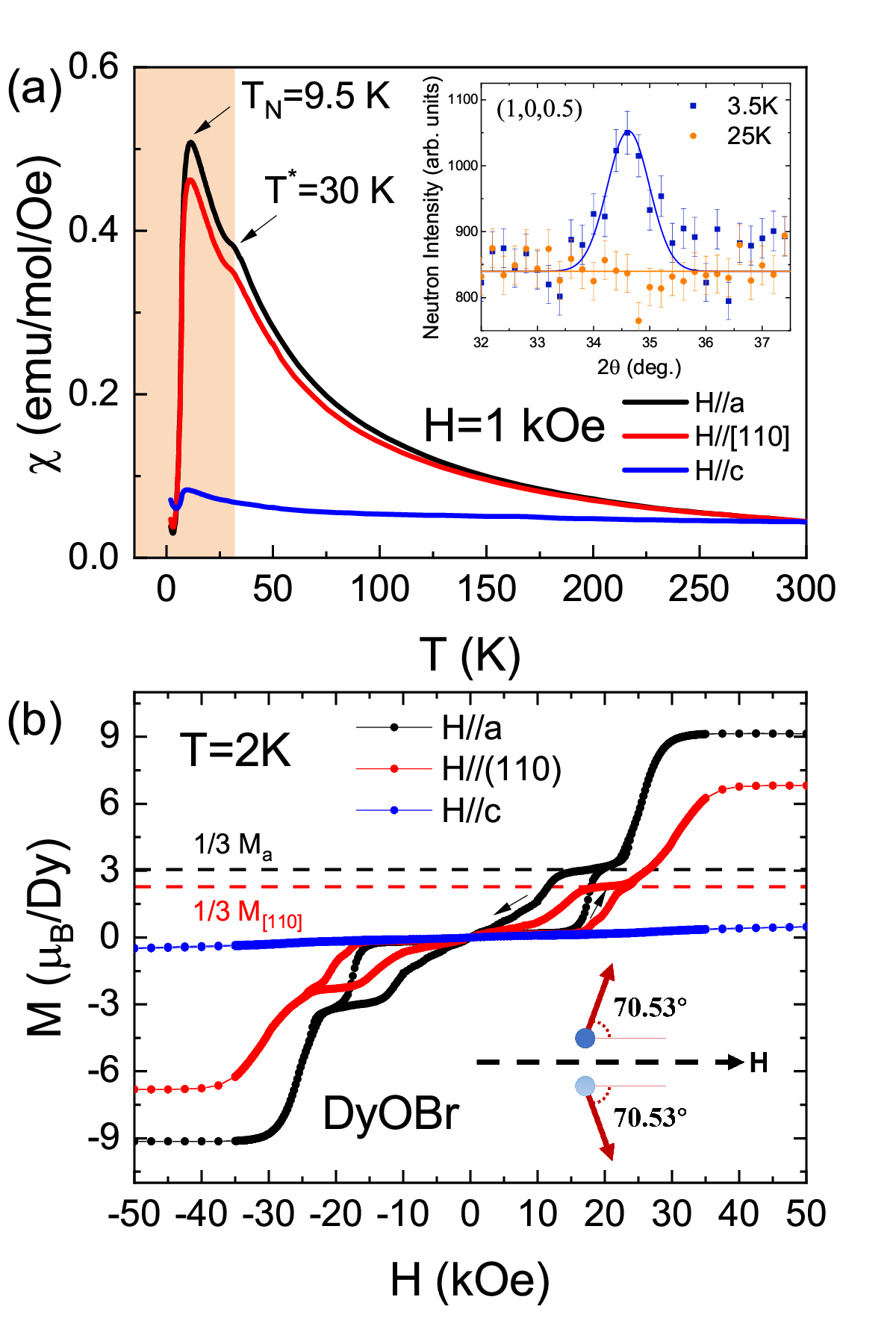}
	\caption {Anisotropic magnetization data for the DyOBr single crystal with magnetic field applied along $a$-, $[110]$- and $c$-axis respectively: (a) The temperature dependent magnetic susceptibilities at $H$=1~kOe. The inset shows the magnetic Bragg peak (1 0 0.5) identified from powder neutron diffraction. (b) Magnetic hysteresis loops at $T$=2~K. The inset show a possible magnetic structure in the 1/3 magnetization plateau.} \label{Fig2}
\end{figure}

The temperature dependent magnetic susceptibility $\chi$(T) of DyOBr is shown in Fig. \ref{Fig2}(a). The Curie-Weiss fit of high temperature paramagnetic susceptibility suggests antiferromagntic interaction exists in DyOBr with details shown in the supplementary material. A sharp onset antiferromagnetic transition appears at T$_N$=9.5~K, which is most prominent for H$\parallel$a, indicating the ordered moment of Dy$^{3+}$ is along $a$-axis. This is consistent with the results of isothermal magnetization presented in Fig. \ref{Fig2}(b). For H$\parallel$a, DyOBr firstly undergoes a spin-flop-like transition at 15~kOe with significant hysteresis behavior, then can be polarized to a ferromagnetic state at 30~kOe with saturation moment of $9.2~\mu_{B}$/Dy, which is quite close to the theoretical value $\mu_J \sqrt{J/(J+1)}$ in a localized model. For H$\parallel$[110], the field-induced magnetic transitions appear at relatively higher field and the saturation moment at 50~kOe is 6.8$\mu_{B}$/Dy$^{3+}$. Meanwhile the magnetization under H$\parallel$c is quite small. The above results confirm DyOBr has a strong uniaxial magnetic anisotropy along $a$-axis, similar to that of DyOCl\cite{DyOCl}. In addition, there is a small upturn in the M(T) curve below 4~K which should be due to some minor ferromagnetic impurity powders since it is more obvious in the powder samples (Fig. S1).

\begin{figure*}[htbp]
	\centering
	\includegraphics[width=\textwidth]{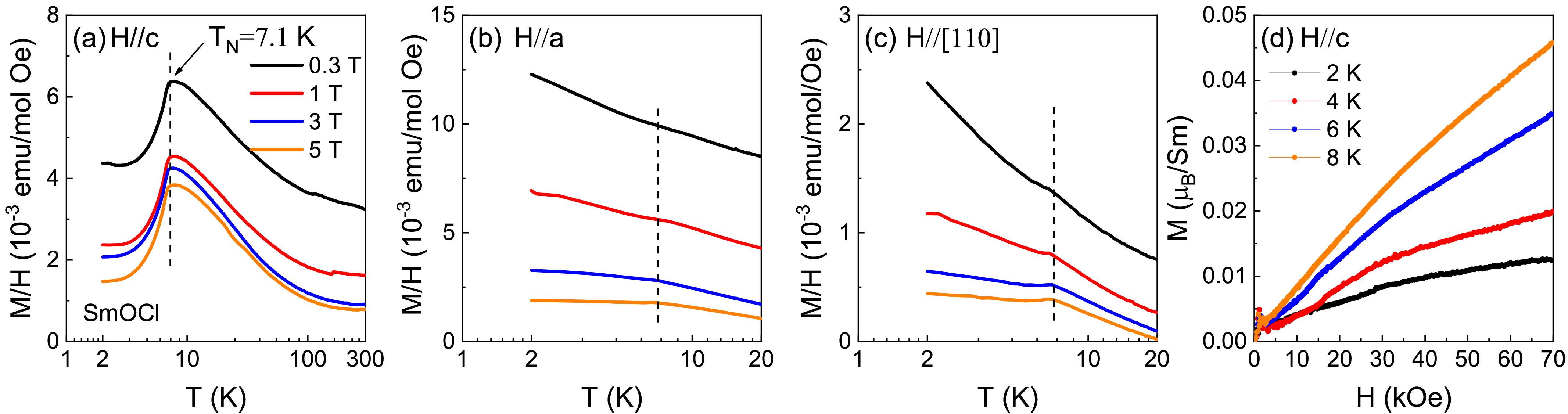}
	\caption {Anisotropic magnetization data for the SmOCl single crystal. The temperature dependent magnetic susceptibilities under H$\parallel$c, H$\parallel$a and H$\parallel$[110] are shown in (a), (b) and (c) respectively. (d) The isothermal magnetization of SmOCl under H$\parallel$c.} \label{Fig3}
\end{figure*}

The spin-flop-like transition for DyOBr displays a clear plateau feature. Under both H$\parallel$a and H$\parallel$[110], DyOBr has a magnetization plateau at about M$_{s}$/3. Previously we found isostructural DyOCl also exhibits a one-third magnetization plateau\cite{DyOCl}. For spin-1/2 antiferromagnets with triangular lattice, 1/3 magnetization plateaus have been observed and widely investigated, which are usually associated with the geometrical frustration and interpreted as arising from an order-by-disorder mechanism driven by quantum fluctuations\cite{p1,p2,p3,p4,p5,p6,p7}. Very recently, the 1/3 magnetization plateau is observed in Na$_3$Ni$_2$BiO$_6$ with honeycomb lattice which results from bond-anisotropic Kitaev interactions.\cite{Wen_plateau} It is quite interesting that for tetrogonal magnetic lattice with Dy$^{3+}$ which has a very large spin could realize a magnetization plateau where the magnetization is also pinned at one-third of its saturation value. One possible explanation from a more classical physical picture is that, due to the magnetic anisotropy, a spin-flop phase with spins from different sublattice have a canting angle of arccos(1/3) forms, which is shown in the inset of Fig. 2(b). On the other hand, this 1/3 plateau may also be a quantum magnetic state such as a magnetic quantum tunneling state from J$_z$=15/2 to J$_z$=5/2. These speculations need to be further confirmed by neutron scattering under magnetic fields and theoretical calculations.

\begin{figure}
	\includegraphics[width=7.5cm]{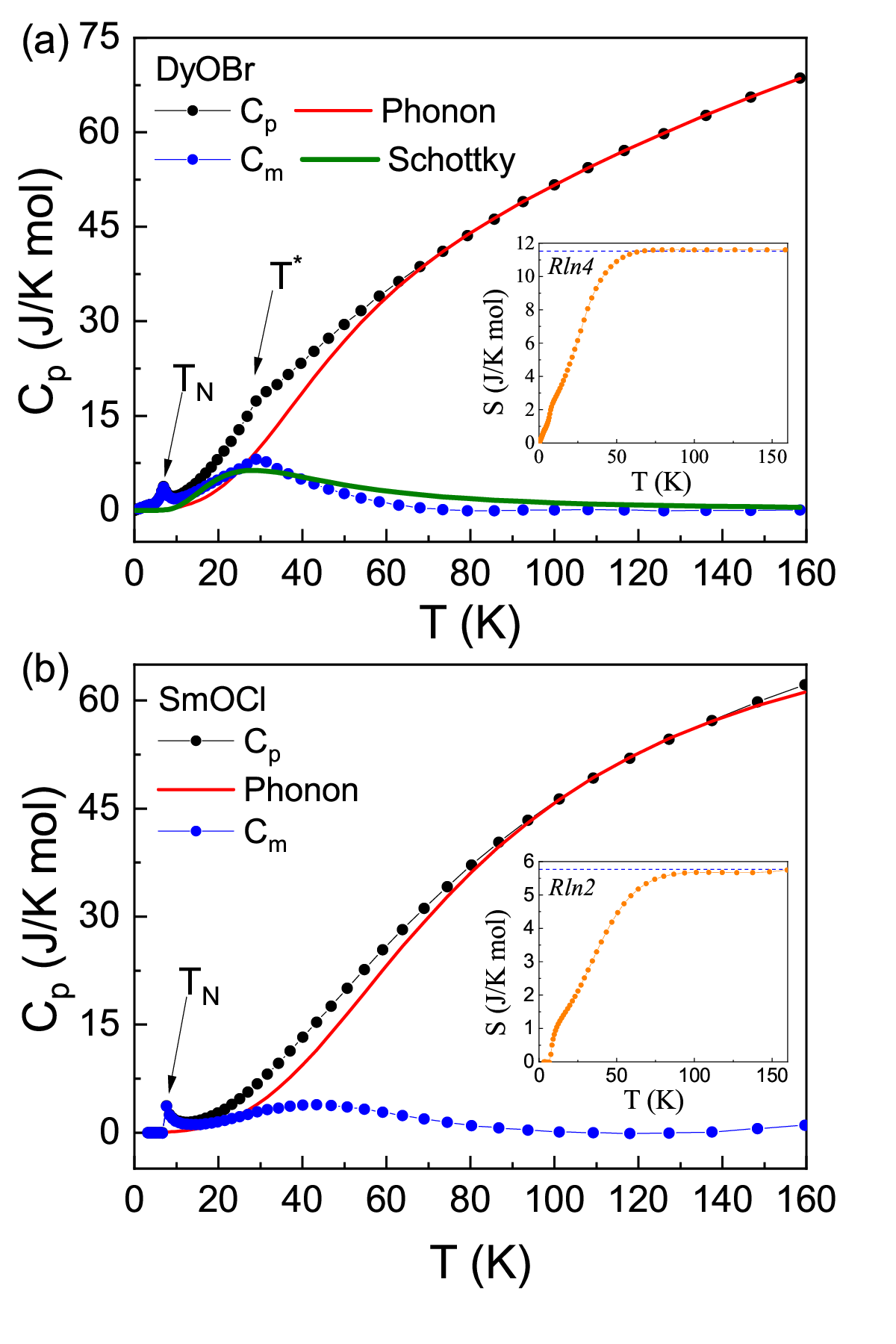}
	\caption {Temperature-dependent specific heat of DyOBr and SmOCl are shown in (a) and (b) respectively. The magnetic specific heat $C_{mag}$ is derived by subtracting phonon contributions through a weighted Debye combined Einstein-model-based curve fit for both DyOBr and SmOCl. Then the calculated magnetic entropy is shown in the insets.} \label{Fig4}
\end{figure}

The long-range antiferromagnetic order in DyOBr could also be confirmed by the clear observation of magnetic Bragg peak (1 0 0.5) below T$_N$ via neutron diffraction, as shown in the inset of Fig. \ref{Fig2}(a). The index of this magnetic peak implies the magnetic unit cell is doubled along the $c$-axis with respect to the crystal unit cell. However the overall neutron data quality could not allow us to solve the accurate magnetic structure. Moreover, the temperature dependent magnetic susceptibility exhibits an anomaly at T$^*$=30~K. This anomaly possibly results from the thermal occupation of low lying crystal field levels as in some other rare-earth compounds\cite{TbV6Sn6} and will be discussed later combined with the heat capacity data.  

The results of magnetization measurements on a SmOCl single crystal are displayed in Fig. \ref{Fig3}. There is a very sharp drop of the susceptibility below T$_N$=7.1~K under H$\parallel$c indicating an antiferromagnetic transition. The isothermal magnetization below T$_N$ shown in Fig. \ref{Fig3}(d) is also quite weak, consistent with an antiferromagnetic ground state. There are some anomalies and nonlinearity in the M(H) curves, which should be due to the disturbance from instrumental background since the crystal has a rather small mass. While for either H$\parallel$a or H$\parallel$[110], the susceptibility anomaly at T$_{N}$ is quite weak in contrast to the sharp cusp under H$\parallel$c. This is a typical feature for antiferromagnet with ordered moments aligned along the $c$-axis. Furthermore, the antiferromagnetic transition is quite robust under field. T$_N$ has nearly zero shift under $\mu_{0}H=5~T$.

The temperature dependent heat capacity data of polycrystalline samples are shown in Fig. \ref{Fig4}. For both DyOBr and SmOCl, C$_p$(T) exhibit a sharp anomaly near T$_{N}$, consistent with the occurrence of antiferromagnetic transition. Since the fabrication of phase-pure LaOBr and LaOCl samples were unsuccessful, we estimated the phonon contribution in C$_p$(T) by a combined Einstein and Debye Model shown as the red solid lines in Fig. \ref{Fig4}. For DyOBr, a shoulder-like feature appears at T$^{*}$=30~K in both C$_p$(T) and magnetic specific heat C$_m$(T). If we consider this anomaly as a Schottky anomaly expected from the thermal population of a low lying crystal field state, then the energy gap could be estimated using a simple two-level model $Rg(\frac{\Delta}{T})^2\frac{e^{\Delta / T}}{[1+ge^{\Delta / T}]^2}$, where $R$ is the ideal gas constant, $g$ is the degeneracy ratio between ground level and excited level, $\Delta$ is the energy gap between two levels. The obtained $\Delta$ from fitting is around $6~meV$. Details about estimating phonon contribution and Schottky fitting are described in the supplementary material. We should mention that the 30~K anomaly in DyOBr may also originate from some unknown magnetic or structural phase transitions. On the other hand, the magnetic entropy could also be estimated based on the specific heat data, it approaches $S=R~ln4$ for DyOBr and $S=R~ln2$ for SmOCl. It should be noted that magnetic entropy below 3~K (the lowest measuring temperature) is not included in this sum.

A previous theoretical calculation on the magnetocrystalline anisotropy of DyOBr suggests it has an easy-plane magnetic anisotropy with moment lying along $a$-axis\cite{DOB_cal}, which is consistent with our experimental result in this report. For rare-earth magnets, the crystalline electrical field splitting of the lowest-lying $4f$ free-ion state becomes a major factor in determining the magnetic anisotropy which is known as the single-ion anisotropy. Therefore, in order to explore the origin of different magnetic anisotropy of DyOBr and SmOCl, we performed crystal field calculations using McPhase based on the point-charge model\cite{McPhase}. We considered the nearest four oxygen neighbors around the rare-earth ion. Since the local point symmetry of the Dy/Sm site is $C_{4v}$ and the crystal $(a,b,c)$ axes are three high-symmetry directions, we used the basis $(x,y,z)$ along the principal crystal axes $(a,b,c)$ respectively. In this local environment, the 16-fold degenerate $J=15/2$ multiplet of Dy$^{3+}$ is split into eight doublet states. The calculated crystal field parameters and wave functions are shown in Table S4-S7 of the supplementary material\cite{Supple}. The calculated ground state wave function suggests the major magnetic component should lie in the $ab$-plane. Based on the calculation result, we can further use McPhase to simulate the low temperature magnetization which also shows that the a-axis is the easy axis which is consistent with the experimental result and previous calculation with other method\cite{DOB_cal}. It is interesting to make a comparison of magnetic anisotropy with another famous vdW magnet CrCl$_3$\cite{CrCl3New}. CrCl$_3$ is also an easy-plane magnet due to the dominate magnetic shape anisotropy which favors an in-plane magnetization. In addition, because of the weak spin-orbital coupling in CrCl$_3$, the in-plane magnetic anisotropy is very weak and in contrast to the uniaxial in-plane anisotropy of DyOBr and DyOCl.

For SmOCl, the 6-fold degenerate $J=5/2$ multiplet of Sm$^{3+}$ is split into three Kramers doublet states under crystal electric field. The calculated ground doublet state is best diagonalized in the basis mentioned above, in which no imaginary coefficients are left in the ground-state wave functions.
The wave function corresponding to the Kramers ground state is 
\[|\psi_{0}\rangle = 0.9968|5/2, \pm 5/2\rangle -0.0806|5/2, \mp 3/2\rangle\]
in $|J,J_z\rangle$ representation. This result clearly demonstrates that the ground-state doublet of Sm$^{3+}$ ion in SmOCl is almost entirely composed of the wave function $|5/2, \pm 5/2\rangle$, with the magnetic easy axis strictly pointing to the $c$-axis (along $z$). This calculated single-ion magnetic anisotropy of SmOCl is in excellent agreement with the experimental magnetization result. Furthermore, the calculated excited crystal field levels are located at energies of $76.34$~meV and $101.29$~meV respectively. This well-separated ground-state doublet indicates that the low-temperature ($T \ll \Delta_{1} \simeq 886 ~K$) magnetic properties are dominated by the ground-state Ising-like magnetic anisotropy. Based on the crystal field calculations, we can further derive the highly anisotropic g-factor as g$_{z}$=1.41 and g$_{xy}$=0.10 for SmOCl.

The diffusion reflectance spectroscopy of DyOBr and SmOCl were measured to determine their band gaps. The data are presented in Fig. 5, where absorption bands are detected at around $235~nm$ and $261~nm$ for DyOBr and SmOCl, respectively. The plot of $[F(R)h\nu]^{0.5}$ versus photo energy $h\nu$ is shown in the figure, where $F(R)=(1-R)^{2}/2R$ is the Kubella-Munk function. Using the methods proposed in \cite{bandgap1, bandgap2}, we calculated the band gaps of DyOBr and SmOCl to be 5.41 eV and 4.74 eV, respectively, indicating that both compounds are insulators.

For both DyOBr and SmOCl, their crystal structures are the same and the difference of lattice constants in the $ab$-plane is quite small. On the other hand, their magnetic anisotropy has contrasting difference. This result may make these two kinds of vdW crystals ideal for designing novel magnetic devices based on vdW heterostructures. Namely one may control the magnetic anisotropy at different layers using different crystals. Therefore, these two compounds reported by us may stimulate further research interests in designing nanoscale magneto-devices. 

\begin{figure}
	\includegraphics[width=7.5cm]{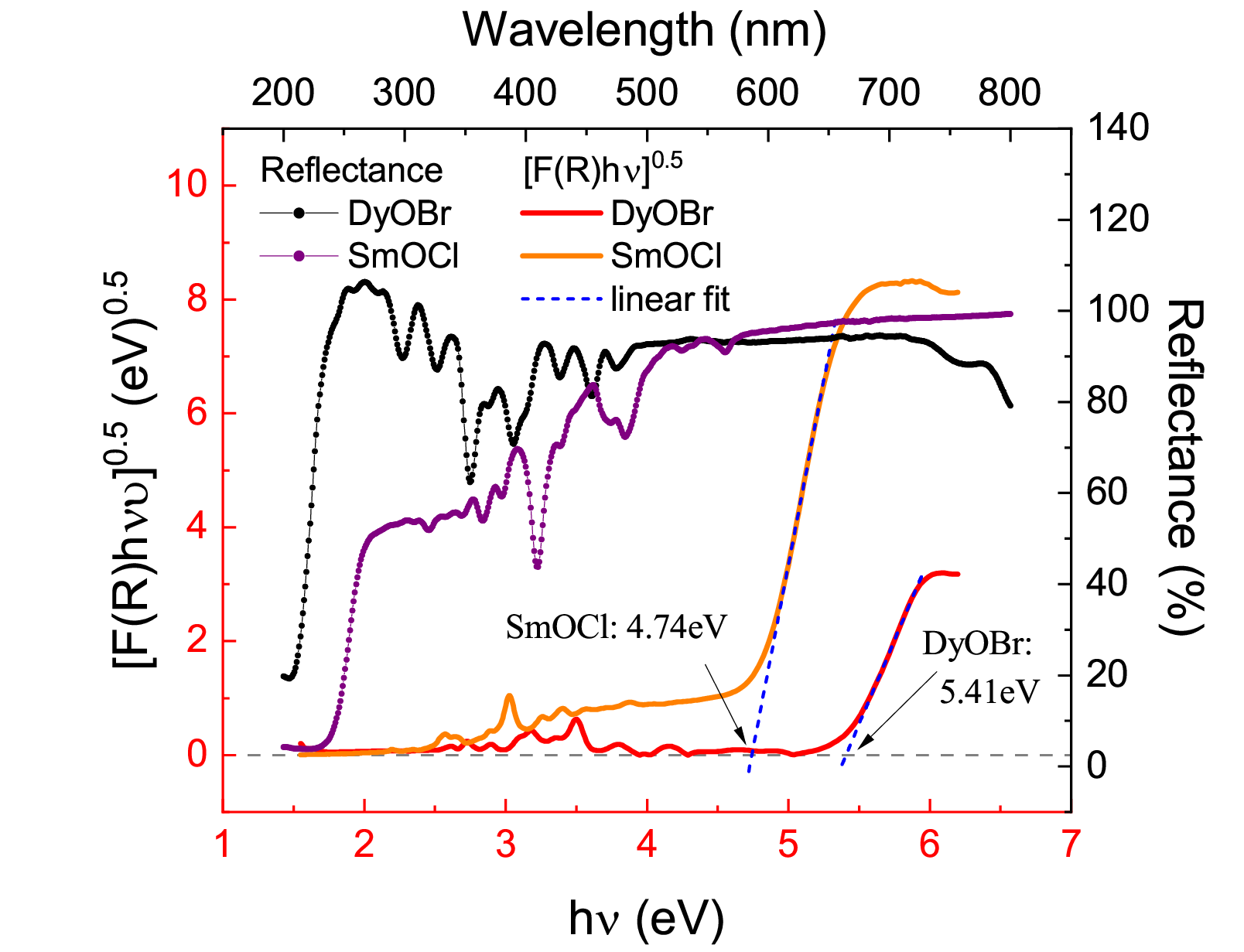}
	\caption {The diffusion reflectance spectroscopy of the DyOBr and SmOCl powder sample (top and right coordinates) and the plot of $[F(R)h\nu]^{0.5}$ versus photoenergy $h\nu$ (bottom and left coordinates).} \label{Fig5}
\end{figure}

\section{Conclusions}

In summary, we have successfully synthesized the single crystals of two van der Waals magnetic insulators DyOBr and SmOCl. DyOBr is an antiferromagnet with a Néel temperature of 9.5 K and a band gap of 5.41 eV. It has a strong magnetic anisotropy along the $a$-axis and susceptibility anomaly at T*=30~K possibly due to the crystal field effect. Besides, DyOBr exhibits a one-third magnetization plateau, suggesting the existence of some exotic field-induced magnetic state. SmOCl is an antiferromagnet with Néel temperature of 7.1~K and a band gap of 4.74~eV. In contrast, SmOCl exhibits strong Ising-like magnetic anisotropy with the easy axis to be $c$-axis, which could be well understood by our crystal field calculations. Both compounds can be exfoliated mechanically into nanoflakes, demonstrating their potential applications in design vdW heterostructures and new-generation spintronic devices.

\section*{Acknowledgement}
This work was supported by the National Natural Science Foundation of China  (No. 12074426, No. 11227906 and No. 12004426), the Fundamental Research Funds for the Central Universities, and the Research Funds of Renmin University of China (Grants No. 22XNKJ40).

\bibliography{DOB}{}
\end{document}